\documentstyle[prd,aps,preprint]{revtex}
\tightenlines

\input epsf.tex
\def\DESepsf(#1 width #2){\epsfxsize=#2 \epsfbox{#1}}

\begin{document}

\draft

\preprint{\vbox{
\hbox{FERMILAB-Pub-00/147-T}
\hbox{OSU-HEP-00-03}\hbox{CTP-TAMU-00-20}
\hbox{UMD-PP-00-084} }}

\title{{\Large\bf Enhanced Electric Dipole Moment of the Muon}\\[0.07in]
{\Large\bf in the Presence of Large Neutrino Mixing}}
\author{{\bf K.S. Babu}$^{1,2}$, {\bf B. Dutta}$^3$ and {\bf R.N. Mohapatra}$^4$ }

\address{$^1$Theory Group, Fermi National Accelerator Laboratory, Batavia, IL 60510, USA \\
$^2$Department of Physics, Oklahoma State University,  Stillwater, OK
74078, USA\footnote{Permanent address}\\
$^3$Center for Theoretical Physics, Department of Physics, Texas A \& M
University, \\ College Station, TX  77843, USA \\
$^4$Department of Physics, University of Maryland, College Park, MD 20742, USA}
\maketitle
\thispagestyle{empty}

\begin{abstract}  The electric dipole moment of the muon ($d^e_{\mu}$) 
is evaluated in supersymmetric models with nonzero neutrino
masses and large neutrino mixing arising from the seesaw mechanism. It is found
that if the seesaw mechanism is embedded in the framework of a left--right
symmetric gauge structure, the interactions responsible for the right--handed
neutrino Majorana masses lead to an enhancement in  $d^e_{\mu}$.  We find $d^e_{\mu}$ as large as
$5\times 10^{-23}$ ecm with a correlated value of $(g-2)_{\mu}\simeq 13\times
10^{-10}$, even for low values of $\tan\beta$. This should provide a strong 
motivation for improving the edm of the
muon to the level of $10^{-24}$ ecm as has recently been proposed.
\end{abstract}

\vskip1.0in

\newpage

\section{Introduction}

It has long been recognized that electric dipole moments (edm) of
fermions can provide 
a unique window to probe into the nature of the forces that are
responsible for CP
violation\cite{barr}. Experimental limits on the edm of neutron have reached the
impressive  level of $6\times 10^{-26}$ ecm \cite{nedm} and have already helped constrain 
and sometimes exclude theoretical models of CP violation. Currently
efforts are under way to improve this limit by
at least two orders of magnitude\cite{lamo}, which will no doubt have very
important implications for physics beyond the standard model. Electric
dipole moment of the electron has severely been constrained 
by atomic measurements in $Cs$
($d^e_e \leq 10^{-26}$) and $T\ell$  ($d^e_e \leq4.3 \times
10^{-27}$ ecm) \cite{commins}.  The limits on the muon edm on the other hand are much
weaker, the
present limit derived from the CERN $(g-2)$ experiment\cite{cern} is
$d^e_\mu \leq 1.1 \times
10^{-18}$ ecm.  There has been a recent proposal to improve this limit on
$d^e_\mu$ to the 
level of $10^{-24}$ ecm \cite{semer}.  In this paper we will argue that
there is a strong
motivation for this proposed improvement, related to the observation of
neutrino masses
and oscillations.  We will show that a natural understanding of  small
neutrino masses
with large oscillation angles in the framework of the seesaw mechanism  will
lead to an enhancement of $d_\mu^e$, to values as large as $5 \times
10^{-23}$ ecm, which is well within the reach
of the proposed experiment.

As for the theory of leptonic edm,  in a large class of models a generic
scaling law holds,
given by  $d^e_{\mu}/d^e_{e}\simeq  m_\mu/m_e$.
If such a relation is valid, even prior to any detailed calculation, one
can infer that the
present upper limit on electron edm will constrain the muon edm to be
less than about 
$10^{-24}$ ecm.  This scaling law arises due to the chiral
structure of the edm operator,
which is very similar to the operator corresponding to the fermion
mass.  To the lowest order in the
light fermion Yukawa couplings, the edm becomes proportional linearly to
the fermion mass.
In specific models, it may so happen that other
constraints put the electron edm itself at a much lower value;  e.g., the
standard
model prediction for the electron edm is $\sim 10^{-41}$ ecm\cite{posp}. 
The scaling law then suggests that the corresponding value for the muon edm
would be at the level of $10^{-39}$ ecm, which is beyond the reach of any
conceivable experiment.  In multi--Higgs doublet extensions of the
standard model,
the dominant contribution to the leptonic edm arises from a two--loop diagram
involving $\gamma-V$--Higgs vertex, where $V=Z,W$ \cite{bz}.  Since such
a vertex is flavor universal,
when converted to the fermion edm, the above--mentioned scaling law will hold.  
Recently an extended Higgs model\cite{barger} has been
analyzed, where it has been shown that for large values of the parameter
$\tan\beta$ (ratio of the
two Higgs vacuum expectation values), the one--loop diagram that scales
as $m_\mu \lambda_\mu^2$,
where $\lambda_\mu$ is the muon Yukawa coupling, can compete with the
two--loop diagram \cite{bz},
leading to order one violation of the scaling law.

In the supersymmetric extension of the standard model (MSSM), under the usual
assumptions about supersymmetry breaking terms, i.e., universality of
scalar mass
terms and proportionality of the trilinear $A$ terms with the
corresponding Yukawa couplings, 
a similar scaling law would hold.  A leading contribution to
leptonic edm in such models is the one--loop diagram involving the bino
virtual state
and a complex $A_{\ell\ell}$ term. The assumption of proportionality of
$A$ terms then
implies that the above mentioned scaling relation remains.  A similar
remark holds when the
chargino diagram is considered, with a complex $\mu$ term, again due to
the universality of
the CP violating parameter. (For a discussion of edm of electrons in MSSM
and SUGRA models, see
ref.\cite{redm3,redm4,redm5,redm6}.) Evaluation of these
bino and  chargino diagrams leads to a value for the muon edm of about
$8 \times 10^{-25}$ ecm, once the upper limit on electron and neutron edm
are satisfied.  The
expected reach of a proposed  BNL experiment for the muon edm is
$10^{-24}$ ecm, which is somewhat above the largest value allowed within the MSSM. 

Recent experimental evidence for neutrino masses, especially from the 
SuperKamiokande atmospheric neutrino
data  \cite{sp}, suggests that the MSSM
must be extended to account for it.  A natural place for small neutrino
masses is the
left--right symmetric extension of the standard model \cite{lr}.  We have
recently advocated a simple
supersymmetric realization of left--right symmetry (SUSYLR) which
accommodates the neutrino masses via the seesaw mechanism\cite{seesaw}.
Our proposal is simply to embed the MSSM into a left--right symmetric
gauge structure at
a high scale $v_R \sim 10^{11}-10^{15}$ GeV.  The effective
MSSM that emerges from this model at scales below the left-right symmetry
breaking scale, $v_R$, is a
constrained MSSM with far fewer number of phases.  In particular,  it has
a built--in solution to the SUSY
CP problem\cite{rasin,bdm1}.  Owing to the constraints of parity
symmetry, the Yukawa coupling matrices
and the trilinear $A$ matrices become hermitian in this
model.  Similarly, the $\mu$ term, the soft $B\mu$ parameter,
and the gluino mass parameters all become real, eliminating potentially
excessive CP violation from the MSSM.  
Furthermore, $R$--Parity arises automatically in this model as part of
the gauge symmetry, since the gauge structure
involves $B-L$ symmetry.

In this paper we wish to investigate the CP violating  muon edm $d^e_\mu$
and $(g-2)_\mu$
in this class of models.  We will show  that the interactions responsible
for the Majorana masses
of the right--handed neutrinos will lead to an enhancement of $d^e_\mu$. 
We find $d^e_\mu$ as large as $5 \times 10^{-23}$ ecm and $(g-2)_\mu$ as much
as $13\times 10^{-10}$.  These values arise even for small  $\tan\beta \sim 3$.
Our main effect 
arises through the renormalization
group extrapolation from the Planck scale to the left--right scale $v_R$
\cite{hall}.  In this interval the
Yukawa couplings of the $\nu_R$ fields which induce their Majorana
masses, as well as the associated trilinear
$A$ terms, will affect the soft supersymmetry breaking parameters of the
effective MSSM, leading to the
enhancement of $d^e_\mu$.  Sine the Majorana Yukawa couplings do not obey $e-\mu$ 
universality, the
scaling law $d_\mu^e/d_e^e = m_\mu/m_e$ is not obeyed by these new diagrams.  

For concreteness, we will work within the framework of a minimal version
of the high scale
SUSYLR (or $SO(10)$) model.  It is minimal in the sense that we
have only one multiplet of Higgs field that gives rise to the usual Dirac
fermion masses, i.e., one
left-right bidoublet $\Phi$ ({\bf 10} in the case of SO(10)). With one such
multiplet, only one Yukawa coupling matrix is allowed in the quark sector,
leading to the proportionality of the up and the down Yukawa coupling
matrices\cite{bdm1,bdm2,ud}. We call this up--down unification.  It has
the consequence that 
all the flavor mixings vanish at the tree level.  We have shown that
acceptable values of the
mixing angles can arise from the one--loop diagrams involving the gluino
(and the chargino), 
proportional to the flavor structure of the trilinear $A$ terms. 
This considerably restricts the flavor and CP violating interactions in
the model and
makes it very predictive. The model has been shown to lead to a consistent
picture of  Kaon CP violation including $\epsilon$ and $\epsilon'$ and it
predicts
neutron edm at the level of $10^{-27}$ ecm.  The leptonic
sector of the model was investigated in Ref.\cite{bdm2}, we shall work
within that
framework to calculate the edm and  $(g-2)$ of the muon.  We have
verified that going to 
non--minimal models, e.g., by employing more than one bidoublet Higgs
field, does not
affect our results by much in the leptonic sector.

\section{Brief overview of the model} 

Let us briefly review the salient features of the 
minimal SUSYLR model.  The electroweak gauge
group of the model is
$SU(2)_L\times SU(2)_R\times U(1)_{B-L}$ with the standard assignment of
quarks and leptons --
left--handed quarks and leptons ($Q,L$) transform as doublets of
$SU(2)_L$, while the right--handed
ones ($Q^c,L^c$) are doublets of $SU(2)_R$.  The Dirac masses of fermions
arise through their Yukawa
couplings to a Higgs bidoublet $\Phi(2,2,0)$.  
The $SU(2)_R\times U(1)_{B-L}$ symmetry is broken to $U(1)_Y$
by $B-L=2$ triplet scalar fields, the left triplet $\Delta$ and right triplet 
$\Delta^c$ (accompanied by $\bar{\Delta}$ and $\bar{\Delta^c}$ fields, their
conjugates to cancel  anomalies).  These fields also couple to the
leptons and are
responsible for inducing large Majorana masses for the $\nu_R$.  An alternative 
is to use $B-L=1$ doublets $\chi $ (left) and
$\chi^c$ (right) along with $\bar{\chi}$ and $\bar{\chi^c}$ instead of
the $\Delta$ fields.  
Here we shall adopt the
$B-L=2$ triplet option, which allows direct couplings to the leptons and
which conserve
$R$--Parity automatically.  Let us write
down the gauge invariant matter part of the superpotential involving these
fields:

\begin{eqnarray}                                                                
W & = &                                                                         
{\bf Y}_q Q^T \tau_2 \Phi \tau_2 Q^c +                                   {\bf
Y}_l L^T \tau_2 \Phi \tau_2 L^c                                    
\nonumber\\                                                                     
  & +  & ( {\bf f} L^T i\tau_2 \Delta L + {\bf f}_c                            
{L^c}^T i\tau_2 \Delta^c L^c)~.
\label{sup}                                                   
\end{eqnarray}  

Under left--right parity, $Q \leftrightarrow Q^{c*}, L \leftrightarrow
L^{c*}, \Phi \leftrightarrow
\Phi^\dagger$, $\Delta\leftrightarrow \Delta^{c*}$, along with
$W_{SU(2)_L} \leftrightarrow W^*_{SU(2)_R}$,
$W_{B-L} \leftrightarrow W^*_{B-L}$ and $\theta \leftrightarrow
\bar{\theta}$.  Here the transformations
apply to the respective superfileds.  As a consequence, ${\bf Y}_q = {\bf
Y}_q^\dagger$, ${\bf Y}_l =
{\bf Y}_l^\dagger$, and ${\bf f} = {\bf f}^*_c$ in Eq. (1).  Furthermore,
the trilinear $A_q$ and
$A_l$ terms will be hermitian, gluino mass term will be real, and the
supersymmetric mass term for $\Phi$
(the $\mu$--term) as well as the supersymmetry breaking $B\mu$ term will
be real.  Departures from these boundary
conditions below $v_R$ due to the renormalization group extrapolation is
small. The model thus provides a natural resolution to the 
supersymmetric CP problem.  

Below $v_R$, the effective theory is the MSSM with the $H_u$ and $H_d$
Higgs multiplets.  
These are contained in the bidoublet $\Phi$ of the SUSYLR model, but in
general they can also
reside partially in other multiplets having identical quantum numbers
under the MSSM symmetry.  
Allowing for such a possibility, the  single coupling matrix ${\bf Y}_q$ of Eq. (1)
describes the flavor mixing in the MSSM in both the up and the down sectors
leading to the relations 
\begin{eqnarray} {\bf Y}_u~=\gamma {\bf Y}_d,
~~~~~~~~~~~~~~~{\bf Y}_{\ell}~=~\gamma~{\bf Y}_{\nu^D},\label{yuk}~
\end{eqnarray} which we call up--down unification. Here $\gamma$ is a
parameter characterizing
how much of $H_u$ and $H_d$ of MSSM are in the bidoublet $\Phi$.  The
case of $H_{u,d}$ entirely in $\Phi$
will correspond to $\gamma = 1$ and $\tan\beta = m_t/m_b$.    
At first sight the first of
the relations in Eq.(2) might appear phenomenologically disastrous since
it leads to
vanishing quark mixings and unacceptable quark mass ratios. We  showed in
Ref.\cite{bdm1} that including the one--loop diagrams involving the
gluino and the chargino
and allowing for a flavor structure for the $A$ terms, 
there exists a large range of parameters (though
not the entire range possible in the usual MSSM) where correct quark mixings as
well as masses can be obtained consistent with flavor changing constraints. 
In Ref.\cite{bdm1}, we explored the parameter
space that allowed for arbitrary squark masses and mixings as well as arbitrary
form for the supersymmetry breaking $A$ matrix. We found a class of solutions
for large $\tan\beta$ $\sim 35-40$ ($\gamma$ =1),  and for small $\tan\beta
\sim 4$ where all quark masses  mixings and CP violating phenomena could be
explained. The smaller value $\tan\beta$ requires larger values of
$\gamma$, since
$\gamma \tan\beta = m_t/m_b$ is fixed.
In this paper, we use small $\tan\beta$ scenarios which is less constrained.

Since the parameter $\gamma$ plays a crucial role in determining the value of
$\tan\beta$, let us explain its origin in an explicit high scale
model.  We will also show
how the solution to the SUSY CP problem can be maintained even for the case of
small $\tan\beta$.  
$\gamma$ arises from the mixing of the bidoublet $\Phi$ with other weak
doublets in 
the high scale theory.  We assume that only one pair of doublets, $H_u$ and $H_d$ of MSSM,
remain light below $v_R$.  
A concrete example which also maintains automatic $R$--Parity of the left--right
model involves the
addition of the following new fields:  $\rho
(2,2,2) + \bar{\rho}(2,2,-2)$ and $\Omega_L (3,1,0) + \Omega_R (1,3,0) $. They
lead to the following new terms in the superpotential:
\begin{eqnarray} W_{\rm new} &=&\mu_\Delta(\Delta \bar{\Delta} + \Delta^c
\bar{\Delta^c})+
\mu_\Phi \Phi^2+
\mu_{\rho} \bar{\rho}\rho + \mu_{\Omega} (\Omega_L^2+\Omega_R^2) \nonumber \\
&+& \lambda_1\left[ {\rm Tr}(\rho \Delta^c \Phi) ~+~ {\rm
Tr}(\bar{\rho}\Delta\Phi)\right] +
\lambda_2 \left[{\rm Tr} (\bar{\rho}\bar{\Delta}^c\Phi) + {\rm Tr}(\rho
\bar{\Delta}\Phi)\right] \nonumber \\
 &+& \lambda_3 {\rm Tr}
(\bar{\Delta}^c\Omega_R \Delta^c ~+~  \bar{\Delta}\Omega_L\Delta)
+\lambda_4 {\rm Tr} (\rho \Omega_R \bar{\rho} ~+~ \bar{\rho} \Omega_L \rho)~.
\end{eqnarray} 
The coupling and the mass parameters in Eq. (3) are guaranteed to be real
by parity symmetry, $P$, defined earlier in combination with the charge
conjugation symmetry $C$ under which all superfields (except $\rho$ and
$\Phi$) transform as $\Psi\rightarrow \Psi^c$, where $\Psi$ stands for a
relevant superfield in the theory; the $W_L\rightarrow W_R$ and
$B\rightarrow -B$. The fields $\rho$ and $\Phi$ transform as
follows: $\rho\rightarrow \tau_2\bar{\rho}^T\tau_2$ and $\Phi\rightarrow
\tau_2\Phi^T\tau_2$. We will assume that the supersymmetry breaking terms
respect only $P$ and not $C$.

It can be shown (see e.g.,
Ref. \cite{melfo}) that this model has a
ground state where $\left\langle \Omega_R\right\rangle 
\sim \left\langle \Delta^c \right\rangle = \left\langle
\overline{\Delta^c}\right\rangle \sim v_R$ and $\left\langle
\Omega_L\right\rangle= \left\langle \Delta \right\rangle = \left\langle \bar{\Delta}\right\rangle =0$. 
The $\rho$ superfield contains an $H_u$--like MSSM doublet and
$\bar{\rho}$ contains an $H_d$--like one. Once the right handed gauge
symmetry is
broken by $\Delta^c$ vev, the doublets in $\Phi$ and those in $\rho$ and
$\bar{\rho}$ mix via a matrix, which is given by
$W_{\rm mass}=\left(\matrix {\rho_u &
\Phi_u}\right)M_{\rm doublet}\left(\matrix {\rho_d \cr
\Phi_d}\right)$, where
\begin{eqnarray} M_{\rm doublet}&=& \left(\matrix{ \mu_{\rho} & \lambda_1
v_R \cr
\lambda_2 v_R & \mu_{\Phi} }\right)~.
\end{eqnarray} 
$M_{\rm doublet}$ being an asymmetric matrix leads to light eigenstates given by
$H_u~=~\cos\theta_1 \Phi_u ~+~ \sin\theta_1 \rho_u $ and $H_d~=~ \cos\theta_2
\Phi_d~+~ \sin\theta_2 {\rho}_d$. Here $\theta_1$ is the $\rho_u-\Phi_u$
mixing angle,
which is unrelated (due to the asymmetry of the matrix) to $\theta_2$, the
$\rho_d-\Phi_d$ mixing angle.  This gives $\gamma
=\frac{\cos\theta_1}{\cos\theta_2}$, which can take any arbitrary value.

We note that due to the combination of $P$ and softly broken $C$ symmetry,
all dimension four couplings are real, leading to a solution to the SUSY
CP problem. To see this, note that due to these symmetries, all entries in
the mass matrix of Eq. (4) are real, so that the
effective $\mu$ term of MSSM stays real.  
(With parity symmetry alone, the $\lambda_{1,2}$
couplings in Eq. (4) could be complex, which would make the effective
$\mu$ term of the MSSM complex.)
Furthermore, since only the dimension 3 and 2 terms of the SUSY
breaking Lagrangian are  assumed to respect P, but not $C$,
such a scenario is completely stable under
renormalization.  (This scheme is distinct from scenarios where CP
symmetry is imposed on
the MSSM Lagrangian at a high scale to solve the SUSY CP problem  
\cite{bb}.  Since the gauge structure
of MSSM does not have parity symmetry, the phases of the soft SUSY  
breaking terms will have to be small in that case.) 

Unlike the large $\tan\beta$ case (corresponding to $\gamma =1$), we are finding
that CP violation in the quark sector has to arise from soft terms.  We
have analyzed this possibility in Ref. \cite{bdm1} and shown its
consistency.
We are pursuing this possibility  further \cite{bdmnew}.  An immediate
outcome
of this scenario for hadronic CP violation is that although there is KM type
CP violation, generically it tends to be sub--leading to SUSY CP violation.  

In the absence of the $\Omega_R$ field in Eq. (3), the doubly charged field $\Delta^{c++}$
in $\Delta^c$ (as well as $\Delta^{c--}$ in $\overline{\Delta^c}$) 
will remain massless -- it will pick up mass only of order the
weak scale, or of order $v_R^2/M_{\rm string}$, if non--renormalizable operators
are included.  Inclusion of $\Omega_R$ (and its left--handed partner $\Omega_L$)
lifts the mass of $\Delta^{c++}$ to the scale $v_R$ \cite{melfo}.  
We will analyze two cases, one with the inclusion of $\Omega_{L,R}$ fields, and one without.
In the latter case, we will take the mass of $\Delta^{c++}$ to be $\sim v_R^2/M_{\rm string}$.

\section{Leptonic CP violation and muon EDM} 

 To discuss CP violation in the lepton sector, we need to specify the
leptonic superpotential
$W_{\ell}$ and the most general soft breaking Lagrangian, ${\cal L}^{\ell}_{\rm soft}$,
in the lepton superpartners.  The leptonic $W_\ell$ is given
in Eq. (1), $\cal{L}_{\rm soft}^\ell$
is given by:
\begin{equation} 
-{\cal L}_{\rm soft}^\ell ~=~ {\bf m}^2_{LL} \tilde{L}^{\dagger} \tilde{L} ~+~
{\bf m}^{2}_{RR}\tilde{L^c}^{\dagger}\tilde{L^c} +\left[A_{l}
\tilde{L}\Phi \tilde{L^c} ~+~ A_{f}(\tilde{L}\tilde{L}\Delta
~+~\tilde{L^c}\tilde{L^c}\Delta^c)~+~ H.c.\right]
\end{equation} 
To generate a nonvanishing muon edm, one needs a complex valued
$ (A_l)_{22}$ and/or complex soft mass-squared terms. But above the scale
where the parity
symmetry is valid, $A_{l}$ is hermitian and therefore its diagonal
elements are all
real. This element can however be complex due to radiative corrections below the
parity breaking scale. There are two ways this can happen: (i) if only parity
symmetry is broken but gauge symmetry $SU(2)_L\times SU(2)_R\times U(1)_{B-L}$
is unbroken at the string scale by introduction of parity odd
singlets\cite{cmp}; (ii) if both
parity and the left-right gauge symmetry are broken, but some remnant of
the ${\bf f}$
and $A_{f}$ couplings remain below the $v_R$ scale. This has been shown to
happen in supersymmetric left-right models with minimal field
content\cite{goran}. In the explicit version described in Sec. II, if the
$\Omega_{L,R}$
fields are absent, the $\Delta^{c++}$ field from $\Delta^c$ will have a
mass of order
$v_R^2/M_{\rm string} \sim 10^{12}$ GeV.  So between $M_{\rm string}$ and
$M_{\Delta^{c++}}$,
the effects of ${\bf f}$ and $A_f$ couplings will be felt, and
$(A_{l})_{22}$ can become complex.  
This will also induce flavor violating complex soft mass--squared terms
proportional to
$A_fA_f^{\dagger}$, even if we start with diagonal soft masses at $M_{\rm
string}$.

In case (i), the way $(A_{l})_{ij}$ become complex is as follows. Below the
D-parity (discrete parity)
breaking scale, $M_{\rm string}$, only $\Delta^c$'s (and not
$\Delta$'s) contribute to
renormalization group equations (RGE) describing the evolution of
$(A_{l})_{ij}$  since the $\Delta$'s acquire masses of order $M_{\rm string}$. 
The  RGE are given in the Appendix for this case.  We have, from
Eq. (25) of Appendix,
\begin{eqnarray} {dA_{l}\over dt}&\propto&
{1\over{16\pi^2}}{(4 A_{f}{\bf f}^{\dagger}{\bf Y}_l+2 {\bf f}{\bf f}^{\dagger}A_l)}~.
\end{eqnarray}
The first term on the RHS of Eq. (6) will introduce phase in $A_l$.  Note
that $A_f$ is
not constrained to be hermitian at the string scale by parity symmetry
(unlike $A_l$, which
must be hermitian at $M_{\rm string}$).  
We will allow for complex entries in the $2-3$ block of $A_f$  in our
analysis.  

Below the D--parity breaking scale, the soft mass parameters ${\bf
m}_{LL}^2$ and ${\bf m}_{RR}^2$ will
evolve differently.  In particular, ${\bf m}_{RR}^2$ will feel the
effects of ${\bf f}$ and $A_f$ couplings.
In order to explain the large oscillation angle needed for the
atmospheric neutrino data, we
will find that $f_{23}$ is not much smaller than $f_{33}$.  Thus
$(A_f)_{23}$ is not much
smaller than $(A_f)_{33}$.  Consequently, $({\bf m}_{RR}^2)_{23}$
will become large and complex.  This is the main source of the enhanced
edm of the muon in
the model.  
This qualitative feature becomes more transparent if we examine the RGE
for ${\bf m}_{RR}^2$ (see
Eq. (27) of Appendix).  It has the form:
\begin{eqnarray} {d{\bf m}^2_{RR}   \over dt}&\propto&
{1\over{16\pi^2}} (2 A_fA_f^{\dagger})~.
\end{eqnarray}
It is clear from Eq.(7) how $({\bf m}_{RR}^2)_{23}$ becomes large and complex.  

The dominant contribution to the edm of muon arises from a diagram
which has right and left--handed muon in the
external legs and a lighter stau  inside the loop.  It utilizes the
above--mentioned
$2-3$ mixing which is large and complex.  
For example, the  diagram can have $\mu_L-\tilde\tau_R$ and
$\tilde\tau_L-\mu_R$ vertices along with the stau mass flip inside the
loop or 
it can involve  just the $\mu_L-\tilde\tau_R$ and $\tilde\tau_R-\mu_R$
vertices.  It might be suspected that similar diagrams will also induce large edm
for the electron.
However, in this model, since
$f_{13}$ and $f_{12}$ are much smaller, such contributions are negligible.
Essentially, we have a scenario where $e-\mu$ flavor symmetry is broken
by a large amount
by the ${\bf f}$ and $A_f$ terms.  As a result the scaling law alluded to
in the introduction
does not hold.  If we assume, as we do in our analysis, the existence of phases
only in the  $2-3$ block of the 
$A_f$ matrix, or if $A_f$ has negligible entries in its first row and
column,
no appreciable edm for the electron gets induced due to mixing
effect. Below $v_R$, 
we have only the MSSM field content. Due to the new ${\bf f}$ couplings above the
$v_R$ scale, 
the $\tilde\tau_1$ mass is lower than usual SUGRA model for the same
values of the 
parameter space (i.e., $m_0$, $m_{1/2}$, $A_0$, $\tan\beta$).  This is why
the diagram 
involving the $\tilde{\tau}$ tends to dominate in $d^e_\mu$.

In case (ii), we use the fact that in the minimal SUSYLR model (without
$\Omega_{L,R}$), $\Delta^{c++}$ and
${\Delta}^{c--}$ remain below the $v_R$ scale; therefore their couplings to
the charged fermions via RGE's lead to imaginary parts in
$(A_l)_{22}$ by an amount $\frac{({\bf f}A_{f}^{\dagger}{\bf
Y}_l)_{22}}{16\pi^2}$.  Again the soft
masses become complex in the same fashion as in case (i).
These fields get decoupled at somewhat lower scale
$\sim 10^{11}$ GeV, below which the spectrum is that of MSSM.



\section{Results} 

Let us first discuss the neutrino mass fits in this model. We
start with a basis where the charged leptons masses are diagonal and
Dirac neutrino masses are given by
\begin{equation} M_{\nu^D}=\gamma\tan\beta M_l,
\end{equation} where $M_l=Diag(m_e,m_{\mu}, m_{\tau})$. The light Majorana
neutrino mass  matrix is then given by:
\begin{equation} M_{\nu}={\gamma^2\tan^2\beta\over v_R} M_l{\bf f^{-1}}M_l,
\end{equation} 
where $\bf f$ is the right--handed Majorana Yukawa coupling matrix.

In our fit, we first use the small angle MSW oscillations solution for the solar
neutrino deficit with  $\Delta m^2_{e\mu}\simeq (0.3-1)\times 10^{-5}$ eV$^2$
and $2\times 10^{-3}\leq \sin^22\theta_{e\mu}\leq 2\times 10^{-3}$. We also use
the $\nu_{\mu}\rightarrow\nu_{\tau}$ oscillation scenario to explain the
observed deficit in
the flux of muon neutrinos from the atmosphere \cite{sp}. The mass splitting is
taken to be  $\Delta m^2_{\mu\tau}\simeq (0.1-1)\times 10^{-2}$ eV$^2$ and the
oscillation angle to be  
$\sin^22\theta_{\mu\tau}\simeq 0.8-1$.

For  tan$\beta$ = 3, we find a good fit to the solar and atmospheric
neutrino data by
choosing ${\bf f}$ at $M_{\rm string}$ to be
\begin{eqnarray} {\bf f}&=&\left(\matrix{
  -1.00\times10^{-4}                & 8.8\times 10^{-4}   &-2.2\times 10^{-5}
\cr
  8.8\times 10^{-4}            &-1.3\times 10^{-2}       &1.03\times 10^{-1} \cr
     -2.4\times 10^{-5}          &1.03\times 10^{-1}        &-1.59   }\right).
\end{eqnarray} 
The resulting neutrino masses at $v_R=10^{15.3}$ GeV are:
$(6.27\times 10^{-6},\, 2.5\times 10^{-3},\,  5.2\times 10^{-2})$ eV.  The
leptonic mixing matrix is given by:
\begin{eqnarray} U&=&\left(\matrix{
 -0.99                & 4.2\times 10^{-2}   & -8.4\times 10^{-5}    \cr
  3.1\times 10^{-2}            &0.74      &-0.67\cr
              -2.9\times 10^{-2}               &-0.71       &-0.71   }\right).
\end{eqnarray}
$U_{21}$ is the mixing angle relevant for  solar neutrino oscillations. 
(Our notation is such that $U M_\nu U^T = M_\nu^{\rm diagonal}$.) 
This choice leads to a simultaneous explanation of the solar and atmospheric
neutrino anomalies.  Note
that we have taken
all Yukawa couplings to be real, consistent with our assumption that $C$
and $P$ symmetry are respected by $d=4$ terms.  

It is possible to fit the large angle oscillations solution to satisfy the solar
neutrino deficit. In that case we take ${\bf f}$ matrix is at $v_R\sim
10^{15.6}$ GeV to be
\begin{eqnarray} {\bf f}&=&\left(\matrix{
  -1.77\times10^{-7}                & -1.42\times 10^{-6}   &0
\cr
  -1.42\times 10^{-6}            &-3.9\times 10^{-3}       &-6.4\times
10^{-2} \cr
     0          &-6.4\times 10^{-2}        &-1.28   }\right).
\end{eqnarray}  With these values the  neutrino masses are
$(1.7\times 10^{-3},\, 2.0\times 10^{-3},\,  3.4\times 10^{-2})$ eV and
the corresponding leptonic mixing matrix
is:\begin{eqnarray} U&=&\left(\matrix{
 -0.89                & -3.3\times 10^{-1}   & -4.0\times 10^{-1}    \cr
  -4.5\times 10^{-1}            &0.63      &0.63\cr
              -2.8\times 10^{-2}               &0.72       &-0.69   }\right).
\end{eqnarray}

We use the one--loop  Yukawa
and two--loop gauge RGE to extrapolate all parameters  between the string
scale and the $v_R$ scale.  Since
the new couplings ${\bf f}$ affect the RGE for the leptonic Yukawas,  one
needs to make sure that the charged lepton masses come out to be correct 
 at the
weak scale.  For
simplicity we choose a universal scenario, i.e., all the scalar masses are
given by a common mass parameter $m_0$ at the string scale. We also assume
a common
trilinear mass
$A_0$($\times {\bf Y}_l$)  for all generations. For
$A_f$ we use a structure similar to $\bf f$. But we do not impose $A_f
\propto \bf f$.
We demand electroweak symmetry to be
broken radiatively. In case (i), where parity is broken at  $M_{\rm string}$,
$\Delta$ fields get decoupled and only the  $\Delta^c$ fields contribute to the
RGE for soft masses. Consequently the renormalized right handed slepton
masses get lowered due to the presence of the new couplings $\bf f$. Furthermore,
$A_l$ will pick up off--diagonal elements and will lose its hermitian
structure through
renormalization. The $\Delta^c$ fields get decoupled at the left-right breaking
scale $v_R$, below which we use the RGE corresponding to the  MSSM degrees.

The EDM for a spin $1/2$ fermion is given by the following effective Lagrangian:
\begin{equation} L_f=-{1\over 2}d_f{\bar\psi}\sigma_{\mu\nu}\gamma_5\psi
F^{\mu\nu}~.\end{equation} In this model, we have only the neutralino-slepton 
loop contribution to the edm of muon. This contribution is given as \cite{redm5}:
\begin{equation} d_\mu^e/e={\alpha_{em}\over {4 \pi
sin^2\theta_w}}\sum_{i=1}^6\sum_{i=1}^4Im(\eta_{\mu ik}){\tilde
m_{\chi^0_i}\over
{\tilde m^2_{k}}}Q_\mu B({{\tilde m_{\chi^0_i}}^2\over {\tilde m^2_{k}}})
\end{equation} where $\eta_{\mu ik}=[-{\sqrt
2}(\tan\theta_W(Q_\mu-T_{3\mu})X_{1i}+T_{3\mu}X_{2i})\Gamma^*_{L2k}+x_\mu X_{3i}
\Gamma^*_{R2k}]({\sqrt 2}\tan\theta_W Q_\mu X_{1i}\Gamma_{R2k}-x_\mu X_{3i}
\Gamma^*_{L2k})$ and $x_\mu={m_{\mu}\over{{\sqrt 2} m_W \cos\beta}}$. $X$
diagonalizes the neutralino mass matrix,
$X^T M_{\chi^0}X=diag(m_{\chi^0_1},m_{\chi^0_2},m_{\chi^0_3},m_{\chi^0_14})$.
Here $\Gamma_{L,R}$ are $6\times 3 $ matrices given by 
 $\tilde q_L=\Gamma_{L,R}\tilde q$ and $B(r)={1\over{2(r-1)^2}}(1+r+2r{\rm ln}r/(1+r))$.

We first analyzed the case where $A_f$ and ${\bf f}$ are proportional.  It still
allows for an overall phase in $A_f$, consistent with $P$ invariance.  In this
case $d_\mu^e$ is highly suppressed, $d_\mu^e \le 10^{-26}$ ecm.  The reason is that
with only one matrix structure ${\bf f}$, when the effective $(A_l)_{22}$ is computed
in the original gauge bases, it will remain real.  Small contribution will arise
in the mixed $\mu-\tau$ EDM operator, which can lead to a small value of $d_\mu^e$
since the physical $\mu$ is a linear combination of the two states.  However, this
$\mu-\tau$ mixing turns out to be small.     
As soon as the proportionality $A_f \propto  {\bf f}$ is relaxed, $d_\mu^e$ becomes
much larger.  We have
analyzed the case where $A_f$ and ${\bf f}$ are non--proportional, but
the magnitudes  $|{A_f}|_{ij}$ are proportional to
 $|{\bf f}|_{ij}$. We allow phases of order 1 in the (23), (32) and (33) elements of
${A_f}$ matrix, while keeping $f_{ij}$ real. In this case we find the
maximum muon edm to be $7\times10^{-25}$ ecm. 
When this assumption of proportionality of the magnitudes is relaxed, even larger
value of $d_\mu^e$ results.  We give an explicit example for this case below.
It should be mentioned that
large values of ${A_f}$ reduces  stau mass while it increases $d_\mu^e$. 
So in exploring regions of large $d_\mu^e$, we need to consider the experimental
limits on stau.  In our
calculation we take the lightest stau mass ($\tilde\tau_1$) to be
$\geq80$ GeV (which is above
the current experimental limit of 70 GeV\cite{opal} at $\sqrt s=202$ GeV). 
In Fig. 1, we exhibit the case which has small angle oscillation solution.
The
large angle solution, however, does not show any difference. In Fig. 1 we plot
the muon edm parameter
$k_\mu \equiv Log_{10}[{{d_\mu^e}\over {1\times 10^{-23}{\rm ecm}}}]$ for case
(i) for
tan$\beta=3$. This corresponds to $D$--parity broken at the string scale,
but left--right
gauge symmetry broken at $v_R \simeq 10^{15.3}$ GeV.  
At the string scale (taken to be $10^{17}$ GeV), we have assumed (in GeV
units throughout)
\begin{eqnarray} {A_f}&=&\left(\matrix{
  -2\times10^{-3}                & 1\times 10^{-2}   &0
\cr
  1.0\times 10^{-2}            &-1\times 10^{2} e^{i\pi/2}       &4.7\times
10^{2}e^{i\pi/2}  \cr
     0          &4.7\times 10^{2} e^{i\pi/2}       &3.3\times
10^{2}e^{-i\pi/2}   }\right).
\end{eqnarray}
We put $A_0=-120$ GeV (with $A_l = A_0 {\bf Y}_l$). 
The solid line in Fig. 1 is drawn for $m_0$=160 GeV. The extreme left corner of the  curve
corresponds to lighter stau mass ($\tilde
\tau_1$)=82 GeV. At the same spot in the parameter space,  the lightest chargino
($\chi^\pm_1$) and  the lightest neutralino masses ($\chi^0_1$) are  106 GeV and
52 GeV respectively. We can see that the muon edm can be as large $\sim 3\times
10^{-23}$ ecm in this case. The dotted  line is drawn for $m_0$=170 GeV for the 
same set of
input values.

In Fig. 2 we plot the muon edm parameter
$k_\mu$, for case (ii) with tan$\beta=3$  and
$m_0=160$ GeV. This case corresponds to $\Delta^{c++}$ surviving below
$v_R$.  We assume the
scale at which it decouples to be $10^{12}$ GeV.  
We have used the universal scenario for the slepton masses and 
have used the same $\bf f$ matrix as before.  At the string  scale, we
take (in GeV)  
\begin{eqnarray} {A_f}&=&\left(\matrix{
  -2\times10^{-3}                & 1\times 10^{-2}   &0
\cr
  1.0\times 10^{-2}            &-1\times 10^{1} e^{i\pi/2}       &3.0\times
10^{2}e^{i\pi/2}  \cr
     0          & 3.0\times 10^{2} e^{i\pi/2}       &1.1\times
10^{2}e^{-i\pi/2}  
}\right).
\end{eqnarray}
We take $A_0=0$ GeV. The extreme left corner of the curve in Fig. 2 corresponds to lighter
stau mass  ($\tilde \tau_1$) mass of 80 GeV. At the same spot, as before, 
 the $\chi^\pm_1$ and  the $\chi^0_1$ masses ($\chi^0_1$) are  106 GeV and
52 GeV respectively.  As can be seen from the figure, large values of
$d_\mu^e$ are possible, as large as $5 \times 10^{-23}$ ecm.  

We have assumed non--proportionality of $A_f$ and $\bf f$ in the
preceding two examples.  
We will argue that this is not unnatural.  First of all, there are no
strong experimental
hints that  suggest proportionality of the two (unlike the case of $A_l$
and ${\bf Y}_l$).  
Second, we have proposed recently a model based on horizontal gauge
symmetry which allows
for all parameters of the soft breaking sector to be arbitrary, subject
only to the
constraints of the horizontal symmetry $H$ \cite{bm}.   The symmetry $H$ was taken to be
$SU(2)_H \times U(1)_H$, with the first two generations of fermions falling into
$SU(2)_H$ doublets and the thrid generation into singlets.  The first two
generations
have $U(1)_H$ charges of $-1$, while the third generation is
neutral.  $H$ is spontaneously
broken by a pair of doublet [$\phi(+1),\bar{\phi}(-1)$] and  singlet
[$\chi(+1), \bar{\chi}(-1)$] scalars fields
whose vev's are below the string scale.  We denote $\epsilon_\phi \equiv
\left\langle 
\phi \right\rangle/M_{\rm string}$, $\epsilon_\chi \equiv \left\langle
\chi \right\rangle/M_{\rm
string}$ with $\epsilon_\phi \sim 1/7, \epsilon_\chi  \sim 1/25$.  The effective
Yukawa couplings involving the light two generations will be proportional to 
powers of $\epsilon_\phi$ and $\epsilon_\chi$.  The $U(1)_H$ also
alleviates potential
problems with $D$--terms associated with horizontal symmetries.  

Within the $SU(2)_H \times (U1)_H$ model, it is not necessary to assume
universality of scalar
masses or proportionality of $A$ terms and the Yukawa couplings.  For the
first two generations,
the scalar masses will be approximately equal, owing to the non--Abelian
sector of the horizontal symmetry.
With the horizontal charge assignment given above, we can write down the
most general $H$--symmetric
Yukawa couplings, soft mass terms and $A$ terms.  Since the $A$ terms
become  hierarchical, all
FCNC constraints can be satisfied, even without proportionality
assumption \cite{bm}.

We will now present an example for the muon edm within this horizontal
symmetric framework.
We will embed the model of Ref. \cite{bm} into left--right symmetry at a
high scale.  
Unlike in Ref. \cite{bm}, all the CKM mixing will vanish at tree--level now.  
In a basis where the Yukawa couplings are diagonalized, the Majorana
neutrino coupling
can be written in the following hierarchical form:
\begin{eqnarray} {\bf f}&=&\left(\matrix{
  f_{11}\epsilon_\chi^4/\epsilon_\phi^2                &
f_{12}\epsilon_\chi   &f_{13}\epsilon_\chi^2/\epsilon_\phi
\cr
  f_{12}\epsilon_\chi            &f_{22}\epsilon_\phi^2
&f_{23}\epsilon_\phi  \cr
     f_{13}\epsilon_\chi^2/\epsilon_\phi          &f_{23}\epsilon_\phi
&f_{33}   }\right).
\end{eqnarray}
 The bilinear soft mass matrix and the A matrix are given as:
\begin{eqnarray} m^2_{RR}&=&\left(\matrix{
  m_0^2               & m_0^2 (x_{12})\epsilon_\chi   &m_0^2
(x_{13})\epsilon_\chi^2/\epsilon_\phi
\cr
  m_0^2 (x_{12})\epsilon_\chi            &m_0^2       &m_0^2
(x_{23})\epsilon_\phi   \cr
     m_0^2 (x_{13})\epsilon_\chi^2/\epsilon_\phi          &m_0^2
(x_{23})\epsilon_\phi         &m_{33}^2  
}\right);\,\,\nonumber \\
\nonumber \\
 A&=&A_0\left(\matrix{
  (y_{11})\epsilon_\chi^4/\epsilon_\phi^2                &
(y_{12})\epsilon_\chi  
&(y_{13})\epsilon_\chi^2/\epsilon_\phi
\cr    
  (y_{21})\epsilon_\chi            &(y_{22})\epsilon_\phi^2
&(y_{23})\epsilon_\phi  \cr
     (y_{31})\epsilon_\chi^2/\epsilon_\phi
&(y_{32})\epsilon_\phi        &y_{33}   }\right).
\end{eqnarray}
We also have $m^2_{LL}={m^2_{RR}}$. This structure for $A$ hold for both
$A_l$ and $A_f$ (as well as for $A_q$). At $M_{\rm string}$ we will take
$A_l$ to be hermitian.  
In order to fit the experimental
values of quark and lepton masses we choose $\epsilon_\phi = 1/7$ and
$\epsilon_\chi = 1/25$.
In this new scenario, the muon edm can be
enhanced to $5\times 10^{-23}$. We have taken soft masses for all the Higgs fields
to be 85 GeV. In Fig. 3, we exhibit  the results for $d_\mu^e$ for one such example. 
To generate this plot, the input values we have used at the string scale
are as follows:
\begin{eqnarray} m^2_{RR}&=&85^2\left(\matrix{
  1                &  {1 \over 2}\epsilon_\chi
&{1\over7}\epsilon_\chi^2/\epsilon_\phi
\cr
  {1 \over 2}\epsilon_\chi            &1       &\epsilon_\phi   \cr
     {1\over7}\epsilon_\chi^2/\epsilon_\phi          &\epsilon_\phi
&1.8  
}\right); A_l=30\left(\matrix{
  \epsilon_\chi^4/\epsilon_\phi^2                & {1\over 2}\epsilon_\chi  
&{1\over 3}\epsilon_\chi^2/\epsilon_\phi
\cr
  {1\over 2}\epsilon_\chi            &{1\over 3}\epsilon_\phi^2
&{1\over 3}\epsilon^{i\pi/3}\epsilon_\phi  \cr
     {1\over 3}\epsilon_\chi^2/\epsilon_\phi          &{1\over
3}\epsilon^{-i\pi/3}\epsilon_\phi        &-4   }\right).
\end{eqnarray}   
\begin{eqnarray} {\bf f}&=&\left(\matrix{
  -1.01\times10^{-4}                & 9.0\times 10^{-4}   &-1.4\times 10^{-3}
\cr
  9.0\times 10^{-4}            &-1.2\times 10^{-2}       &1.04\times 10^{-1} \cr
     -1.4\times 10^{-3}          &1.04\times 10^{-1}        &-1.59
}\right);\,\,
\nonumber \\
\nonumber \\
 A_f&=&500\left(\matrix{
  -\epsilon_\chi^4/\epsilon_\phi^2                & {1\over 3}\epsilon_\chi  
&-{1\over 7}\epsilon_\chi^2/\epsilon_\phi
\cr
  {1\over 3}\epsilon_\chi
&-{3}\epsilon^{i\pi/2}\epsilon_\phi^2
&{4}\epsilon^{i\pi/2}\epsilon_\phi  \cr
     -{1\over 7}\epsilon_\chi^2/\epsilon_\phi
&{4}\epsilon^{i\pi/2}\epsilon_\phi        &0.6\epsilon^{-i\pi/2}   }\right).
\end{eqnarray}
Note that we have allowed for all coefficients to be order one,
consistent with the horizontal symmetry.  (This is also
true for the ${\bf f}$ matrix elements.)
The ${(13),\,(31)}$ elements of the $\bf f$ are no longer very small like
our previous example 
because of the symmetry requirement.

The choice of ${\bf f}$ matrix in this case corresponds to the following
light neutrino masses:  
$(6.27\times 10^{-6},\, 2.9\times 10^{-3},\,  4.4\times 10^{-2})$
eV.  The corresponding 
leptonic mixing matrix is:
\begin{eqnarray} U&=&\left(\matrix{
 -0.99                & 4.2\times 10^{-2}   & -3.2\times 10^{-3}    \cr
  3.2\times 10^{-2}            &0.70      &-0.71\cr
              -2.9\times 10^{-2}               &-0.74       &-0.66   }\right).
\end{eqnarray}

Note that in this example $\nu_e-\nu_\mu$ oscillation explains the solar
neutrino data via small angle MSW oscillation.
$\nu_\mu-\nu_\tau$ oscillation explains the atmospheric neutrino
data.  We have found that by varying the order
one couplings slightly, it is also possible to obtain a different
scenario whre $\nu_e-\nu_\tau$ oscillation
is relevant for solar neutrinos, while $\nu_\mu-\nu_\tau$ oscillations
with $m_{\nu_\mu} \ge m_{\nu_\tau}$
explains the atmospheric neutrino data \cite{bm}.  The predictions for $d_\mu^e$ is
not much altetered in such a scenario.

Now we turn to the evaluation of $(g-2)$ of the muon.    
In MSSM, the $(g-2)_{\mu}$ gets contribution from the chargino and neutralino
diagrams \cite{g-21,g-22,wg,g-23}.    The relevant expressions can be found in
Ref. \cite{wg}. In this model we have
contributions from both these loops.  The chargino contribution is somewhat
bigger than the neutralino loop. We find the magnitude of $(g-2)_{\mu}$ to be 
$(6-10)\times 10^{-10}$ 
for the curves in Figs. 1 and 2 and  $(8-13)\times 10^{-10}$ for the model
with horizontal symmetry given in Fig. 3.
 
As for other rare processes, 
the branching ratio of $\tau\rightarrow\mu\gamma$ is  one to two orders
of magnitude below the present experimental limit. Since this process
cnnot be made much smaller,
it will be of great interest to improve the present limit by two orders of
magnitude, which does not
appear to be out of question.  
In all cases that we studied, the edm for electron is of order $
10^{-28}$ ecm.  As for 
$\mu \rightarrow e \gamma$, it is three to four orders of magnitude
smaller than current limits
for cases (i) and (ii), and one order of magnitude smaller than current
limits in the case
of horizontal symmetry.

In conclusion, we have shown that in supersymmetric extensions of the
stanadard model that accommodates small neutrino masses  via
the seesaw mechanism, there is an enhancement of the muon electric dipole
moment.  Interactions responsible for the generation of Majoran masses
for the right--handed neutrinos are responsible for this enhancemnt
through renormalization group effects.
We have found values of $d_\mu^e$ as large as 
$5 \times 10^{-23}$ ecm.  Our finding should provide a strong motivation
to improve the limit of $d_\mu^e$ to the level of $10^{-24}$ ecm, as has
recently been proposed.  Probing $d_\mu^e$ at this level could reveal
the underlying structure responsible for CP violation as well as for
the generation of neutrino masses.  

\section{Acknowledgments}

One of the authors (R.N.M) would like to thank K. Jungman for discussions on
the present ideas for the searches for muon edm. K.S.B is thankful to the Theory
Group at Fermilab where part of this work was done for its warm hospitality.  
The work of K.S.B is supported by Department of Energy Grant No. DE-FG03-98ER41076
and by a grant from the Research Corporation.
B.D is supported by National Science Foundation Grant No. PHY-9722090.
R.N.M is supported by NSF Grant No. PHY-9802551.

\newpage

\section{Appendix}

In this Appendix we  give the renormalization group equations appropriate for the momentum
range between $M_{\rm string}$ and $v_R$ for the case where parity is broken at $M_{\rm string}$. 
\begin{eqnarray} 
{d {\bf f}\over dt}&=&{1\over {16 \pi^2}} [-4 \pi (7 \alpha_R
    +{9\over2} \alpha_{B-L}) {\bf 1}
    +2 {\bf f}\,{\bf f}^{\dag} +
    4 {\bf Y}_l{\bf Y}_l^{\dag}
    + {\rm Tr}({\bf f}\,{\bf f}^{\dag})]{\bf f},
 \\
{d Y_l\over dt}&=&{1\over {16 \pi^2}} [-4 \pi (3 \alpha_R+3
\alpha_L
    +{3\over2} \alpha_{B-L}) {\bf 1}
    +2 {\bf f}\,{\bf f}^{\dag} +
    4 {\bf Y}_l{\bf Y}_l^{\dag}+{\rm Tr}(3 {\bf Y}_q{{\bf Y}_q}^{\dag}
    + {{\bf Y}_l}{{\bf Y}_l}^{\dag})]{{\bf Y}_l},
\\
{d A_{l}\over dt}&=&{1\over {16 \pi^2}}[-4\pi(3 \alpha_{R} +3
\alpha_{L} +{3\over 2}\alpha_{B-L} )A_{l}\\\nonumber&+&8\pi(3 \alpha_{R}
M_{R}+3 \alpha_{L} M_{L}+{3\over 2}\alpha_{B-L} M_{B-L}){\bf Y}_{l}\\\nonumber&+& 4
A_{l}{\bf Y}_l^{\dag}{\bf Y}_l +
    8 {\bf Y}_l{\bf Y}_l^{\dag} A_{l}+
    2 {\bf f}\,{\bf f}^{\dag}A_{l}+
    4 A_{f}{\bf f}^{\dag}{{\bf Y}_l}\\\nonumber&+&
    2 {\rm Tr}(A_{l}
    {\bf Y}_l^{\dag}){\bf Y}_l+
    {\rm Tr}({\bf Y}_l {\bf Y}_l^{\dag})A_{l} +
    6 {\rm Tr}(A_{q}{\bf Y}_q^{\dag}) {\bf Y}_l +
    3 {\rm Tr}({\bf Y}_q {\bf Y}_q^{\dag}){\bf Y}_l],
\\
{d A_{f}\over dt}&=&{1\over {16 \pi^2}}[-4\pi(7 \alpha_{R}
+{9\over 2}\alpha_{B-L})A_{f}+8\pi(7 \alpha_{R}M_{R} +{9\over
2}\alpha_{B-L}M_{B-L} ){\bf f}\\\nonumber&+&
    8 A_{l}{\bf Y}_l^{\dag}{\bf f} +
        4 {\bf Y}_l{\bf Y}_l^{\dag} A_{f}+
    2 A_{f}{\bf f}^{\dag}{\bf f}+
    4 {\bf f}\,{\bf f}^{\dag}A_{f}\\\nonumber&+&2 {\rm Tr}(A_{f}
    {\bf f}^{\dag}){\bf f}+
    {\rm Tr}({\bf f}\,{\bf f}^{\dag})A_{f}],
\\
{d{\bf m}^2_{RR}\over dt}&=&{2\over {16 \pi^2}}[-4\pi(3/2
\alpha_{B-L} M_{B-L}^2+3 \alpha_{L}
M_{L}^2)\\\nonumber&+&{1\over2}(({\bf Y}_l{\bf Y}_l^{\dag}+{\bf f}\,{\bf f}^{\dag}){\bf
m}^2_{RR}+{\bf m}^2_{RR}({\bf Y}_l{\bf Y}_l^{\dag}+{\bf f}\,{\bf f}^{\dag})+ 2({\bf Y}_l{\bf
m}^2_{RR}{\bf Y}_l^{\dag}) \\\nonumber&+& 2({\bf f}{\bf m}^2_{RR}{\bf f}^{\dag} +
m^2_{\Phi}{\bf Y}_l{\bf Y}_l^{\dag}+m^2_{\Delta^c}{\bf f}\,{\bf
f}^{\dag}+A_{l}A_{l}^{\dag}+A_{f}A_{f}^{\dag}))].
\end{eqnarray}

\begin{figure}[htb]
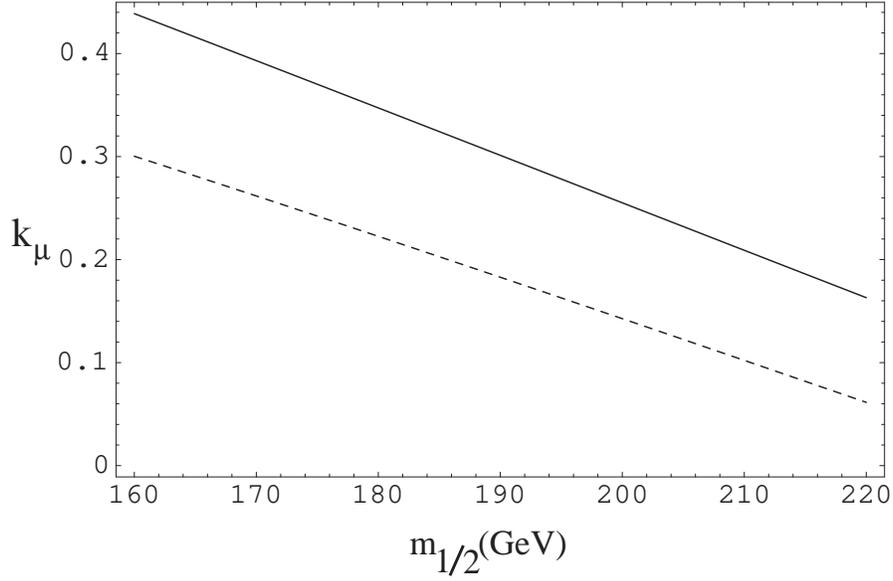

\centerline{ \DESepsf(bdmedm33.epsf width 12 cm) }
\smallskip
\caption {$k_\mu(\equiv Log_{10}[{{d_\mu^e}\over {1\times 10^{-23}{\rm ecm}}}])$ 
is plotted against $m_{1/2}$ for tan$\beta$=3 for case (i). The solid line is for $m_0=160$ GeV 
and the dotted line is for $m_0=170$. The other inputs are described in the text.}
\vspace{1 cm}\end{figure}
\begin{figure}[htb]
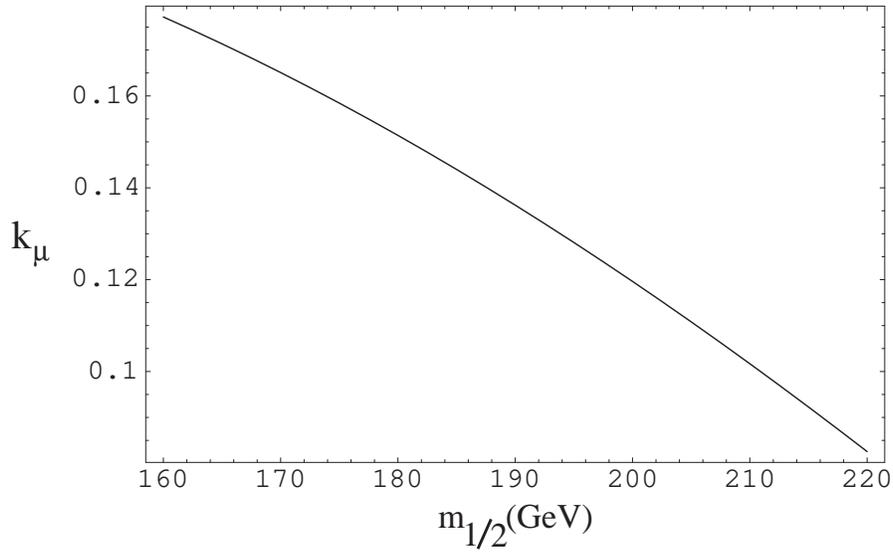

\centerline{ \DESepsf(bdmedm44.epsf width 12 cm) }
\smallskip
\caption {$k_\mu(\equiv Log_{10}[{{d_\mu^e}\over {1\times 10^{-23}{\rm ecm}}}])$ 
is plotted against $m_{1/2}$ for tan$\beta$=3 for case (ii). $m_0$ is 160 GeV.
 The other inputs are described in the text.}
\vspace{0 cm}
\end{figure}

\begin{figure}[htb]
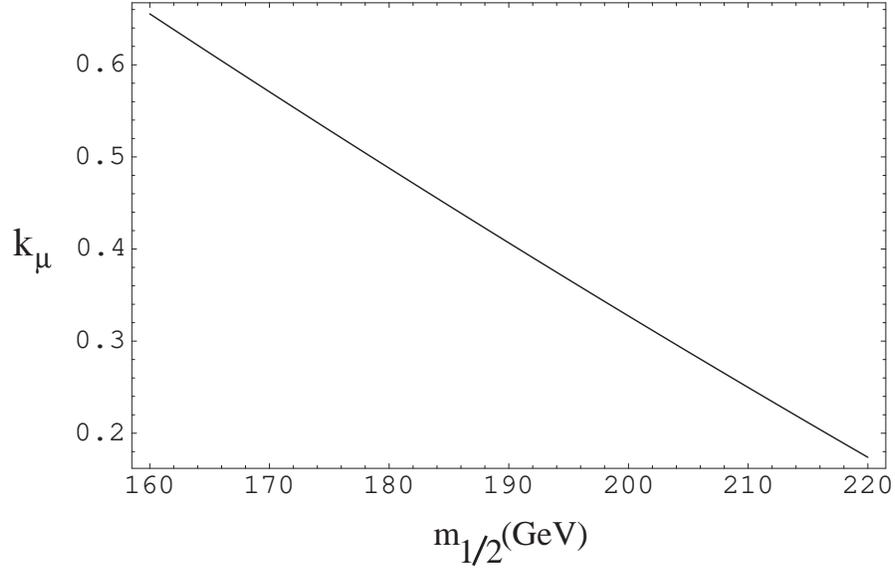

\centerline{ \DESepsf(bdmedm55.epsf width 12 cm) }
\smallskip
\caption {$k_\mu(\equiv Log_{10}[{{d_\mu^e}\over {1\times 10^{-23}{\rm ecm}}}])$ 
is plotted against $m_{1/2}$ for tan$\beta$=3 for the model with horizontal symmetry.
 $m_0$ is 85 GeV.
 The other inputs are described in the text.}
\vspace{0 cm}
\end{figure}

\end{document}